\ifx\documentclass\undefined
\documentstyle[12pt]{article}
\else
\documentclass[12pt]{article}
\fi

\makeatletter
\renewcommand{\theequation}{\thesection.\arabic{equation}}
\@addtoreset{equation}{section}
\def\eqnarray{%
\stepcounter{equation}%
\let\@currentlabel=\theequation
\global\@eqnswtrue
\global\@eqcnt\z@
\tabskip\@centering
\let\\=\@eqncr
$$\halign to \displaywidth\bgroup\@eqnsel\hskip\@centering
$\displaystyle\tabskip\z@{##}$&\global\@eqcnt\@ne
\hfil$\displaystyle{{}##{}}$\hfil
&\global\@eqcnt\tw@$\displaystyle\tabskip\z@{##}$\hfil
\tabskip\@centering&\llap{##}\tabskip\z@\cr}
\makeatother

\def\bbbz{{\mathchoice {\hbox{$\sf\textstyle Z\kern-0.4em Z$}}
{\hbox{$\sf\textstyle Z\kern-0.4em Z$}}
{\hbox{$\sf\scriptstyle Z\kern-0.3em Z$}}
{\hbox{$\sf\scriptscriptstyle Z\kern-0.2em Z$}}}}
\def\bbbq{{\mathchoice {\setbox0=\hbox{$\displaystyle\rm Q$}\hbox{\raise
0.15\ht0\hbox to0pt{\kern0.4\wd0\vrule height0.8\ht0\hss}\box0}}
{\setbox0=\hbox{$\textstyle\rm Q$}\hbox{\raise
0.15\ht0\hbox to0pt{\kern0.4\wd0\vrule height0.8\ht0\hss}\box0}}
{\setbox0=\hbox{$\scriptstyle\rm Q$}\hbox{\raise
0.15\ht0\hbox to0pt{\kern0.4\wd0\vrule height0.7\ht0\hss}\box0}}
{\setbox0=\hbox{$\scriptscriptstyle\rm Q$}\hbox{\raise
0.15\ht0\hbox to0pt{\kern0.4\wd0\vrule height0.7\ht0\hss}\box0}}}}
\def\bbbc{{\mathchoice {\setbox0=\hbox{$\displaystyle \rm C$}\hbox{\raise
0.06\ht0\hbox to0pt{\kern0.4\wd0\vrule height0.9\ht0\hss}\box0}}
{\setbox0=\hbox{$\textstyle\rm C$}\hbox{\raise
0.06\ht0\hbox to0pt{\kern0.4\wd0\vrule height0.9\ht0\hss}\box0}}
{\setbox0=\hbox{$\scriptstyle\rm C$}\hbox{\raise
0.06\ht0\hbox to0pt{\kern0.4\wd0\vrule height0.8\ht0\hss}\box0}}
{\setbox0=\hbox{$\scriptscriptstyle\rm C$}\hbox{\raise
0.06\ht0\hbox to0pt{\kern0.4\wd0\vrule height0.8\ht0\hss}\box0}}}}
\makeatletter
  \renewcommand{\theequation}{%
 \thesection.\arabic{equation}}
  \@addtoreset{equation}{section}
\makeatother
\newtheorem{theorem}{Theorem}[section]
\newtheorem{lemma}{Lemma}[section]

\newtheorem{remark}{Remark}[section]

\newsavebox{\toy}
\savebox{\toy}{\framebox[0.65em]{\rule{0cm}{1ex}}}
\newcommand{\QED}{\usebox{\toy}}
\def\nlni{\par\ifvmode\removelastskip\fi\vskip\baselineskip\noindent}
\newenvironment{proof}{\nlni\begingroup\it Proof.\rm}{
\endgroup\vskip\baselineskip}

\begin{document}
\setlength{\baselineskip}{15pt}
\title{
The repulsion between localization centers 
in  the Anderson model
}
\author{Fumihiko Nakano
\thanks{Faculty of Science, 
Department of Mathematics and Information Science,
Kochi University,
2-5-1, Akebonomachi, Kochi, 780-8520, Japan.
e-mail : 
nakano@math.kochi-u.ac.jp}}
\date{}
\maketitle
\begin{abstract}
In this note
we show that, a simple combination of deep results in the theory of random Schr\"odinger operators yields a quantitative estimate of the fact that the localization centers become far apart, as corresponding energies are close together. 
\end{abstract}


\section{Introduction}
In this paper, 
we consider a simple random system and its spectral region where the Anderson localization holds (i.e., we have dense point spectrum with exponentially decaying eigenfunctions). 
We study 
the ``center" of these localized eigenfunctions and prove a repulsive property on the distribution of those in relation to their corresponding energies, that is, 
``as eigenvalues get closer, the corresponding localization centers become far apart"
$\cdots (*)$.
A naive explanation 
which supports the observation $(*)$ is : 

(1) 
Suppose we have 
two eigenvalues 
$E_1, E_2$
with their corresponding localization centers 
$x_1, x_2$
satisfy 
$| x_1 - x_2 | \sim L$. 
Because 
eigenfunctions are exponentially localized, 
we can find a finite box 
$\Lambda$ 
of size in the order of
$L$ 
surrounding
$x_1, x_2$, 
so that 
$H_{\Lambda}$
has two eigenvalues close to 
$E_1, E_2$. 
If the density of state is finite, 
eigenvalues of 
$H_{\Lambda}$ 
would arrange in the order of 
$| \Lambda |^{-1}$
so that we could have
$| E_1 - E_2 | \ge | \Lambda |^{-1} \sim L^{-d}$.

(2)
(Carmona-Lacroix \cite[p.338]{CL})
Because locally we have no repulsion between eigenvalues(Molchanov, Minami \cite{Molchanov, Minami}), 
the overlap between the eigenfunctions should be small as the corresponding energies get closer. 

The phenomenon $(*)$
has been observed in numerical calculations
\cite[p.338]{CL}.
\cite{Molchanov}
proves $(*)$ in a special model on one space dimension, with a complicated statement. 
On the other hand, the observation 
$(*)$
was used by Mott in the study of the ac-conductivity of random systems whose mathematical study is done recently
(Kirsch-Lenoble-Pastur, Klein-Lenoble-M\"uller \cite{KLP, KLM}).
The purpose of this paper 
is to obtain a quantitative statement of $(*)$ which holds for almost surely. 
To study the same property 
for the averaged quantity would require more elaborate analysis as is done in \cite{KLM}. 

Our model 
is the standard tight binding Hamiltonian with random potential on 
${\bf Z}^d$
as was treated by 
\cite{Minami}.
\[
(H \varphi)(x)
:=
\sum_{|y-x| =1} \varphi (y)
+
\lambda V_{\omega} \varphi (x), 
\quad
\varphi \in l^2 ({\bf Z}^d)
\]
where
$\lambda > 0$
is the coupling constant and 
$\{ V_{\omega} (x) \}_{x \in {\bf Z}^d}$
is independent, identically distributed 
real-valued random variables on a probability space
$(\Omega, {\cal F}, {\bf P})$
whose common distribution is assumed to have a bounded  density 
$\rho$. 
Under this assumption, the following facts are well-known.
(1)
$\sigma (H) = \Sigma := [-2d, 2d] + \mbox{supp } \rho$, a.s.
(Kunz-Souillard \cite{Kunz-Souillard}),
(2)
We can find a bounded interval
$I (\subset \Sigma)$
such that the spectrum of 
$H$ 
in 
$I$
is almost surely pure point with exponentially decaying eigenfunctions
(Anderson localization). 
$I$
can be taken, for instance
(Fr\"olich-Spencer, von Dreifus-Klein, Aizenman \cite{FS, vonDK, Aizenman}), 
(i)
(high disorder) 
$I = \Sigma$ 
if 
$\lambda \gg 1$, 
(ii)
(extreme energy) away from the origin, 
(iii)
(weak disorder) 
away from the spectrum of the free Laplacian if 
$| \lambda | \ll 1$, and 
(iv) band edges. 
Before stating our results, 
we define some notations. 
\\

\noindent
{\bf Definition}\\
{\it 
(1)
$\Lambda_L (x) = 
\{ y \in {\bf Z}^d : | y_j - x_j | \le \frac L2, 
j=1, 2, \cdots, d \}$
is the finite box in 
${\bf Z}^d$
of length 
$L > 0$
with its center 
$x = (x_1, \cdots, x_d) \in {\bf Z}^d$.
For simplicity, 
$\Lambda_L := \Lambda_L (0)$. 
$\partial \Lambda := 
\{ y \in \Lambda : \exists z \notin \Lambda, 
|y - z | = 1\}$
be the boundary of the box 
$\Lambda$. 
$H_{\Lambda}$
is the restriction of 
$H$
on 
$\Lambda$
with Dirichlet boundary condition. 
$G_{\Lambda}(E; x,y) :=
\langle 
\delta_x, (H_{\Lambda}- E)^{-1} \delta_y
\rangle_{l^2({\bf Z}^d)}$
is the matrix element of the resolvent 
$(H_{\Lambda}- E)^{-1}$
where
$\delta_x(z) = 1 (z=x), 0 (z\ne x)$
and 
$\langle \cdot, \cdot \rangle_{l^2({\bf Z}^d)}$
is the inner-product in 
$l^2({\bf Z}^d)$.
$|\Lambda| = \sharp \Lambda$
is the volume of a box 
$\Lambda (\subset {\bf Z}^d)$
and 
$| I | = b - a$
is the width of an interval 
$I = (a,b) (\subset {\bf R})$. 
$\chi_{\Lambda}$
is the characteristic function of a box 
$\Lambda$. 
\\
(2)
We say the box
$\Lambda_L (x)$
is
$(\gamma, E)$-regular
iff 
$E \notin \sigma (H_{\Lambda_L (x)})$
and for any
$y \in \partial \Lambda_L (x)$, 
\[
| G_{\Lambda_L (x)} (E; x, y) | 
\le
e^{- \gamma L/2}.
\]
(3)
For 
$\phi \in l^2 ({\bf Z}^d)$, 
let 
$X({\phi})$
be the set of its localization centers given by
\[
X({\phi}) := 
\left\{
x \in {\bf Z}^d : 
| \phi (x) | = \max_{y \in {\bf Z}^d} | \phi (y) |
\right\}.
\]
This notion 
is due to Germinet-De Bi\`evre \cite{BG}.
Since
$\phi \in l^2 ({\bf Z}^d)$, 
$X({\phi})$
is a finite set. 
Moreover, 
for the set 
$\{ E_j (\omega) \}_{j \ge 1}$
of eigenvalues of 
$H$
counting multiplicity, we take 
the  corresponding eigenfunctions
$\{ \phi_j (\omega) \}_{j \ge 1}$
and let 
$X(E_j(\omega)) := X(\phi_j  (\omega))$.
\\
(4)
For 
a finite box
$\Lambda$
and 
$\phi \in l^2({\bf Z}^d)$, 
we say 
$\phi$
is localized in 
$\Lambda$
iff
$X({\phi}) \cap \Lambda \ne \emptyset$.
For an eigenvalue 
$E_j (\omega)$
of 
$H$, 
we say 
$E_j (\omega)$
is localized in 
$\Lambda$
iff
$X(E_j(\omega)) \cap \Lambda \ne \emptyset$}. 
\\

Let 
$I (\subset \Sigma)$
be the bounded interval where the initial length scale estimate of the multiscale analysis holds : \\
%

\noindent
{\bf Assumption}
\\
{\it 
We have an interval 
$I (\subset \Sigma)$
with 
\[
{\bf P}\left(
\mbox{for $\forall E \in I$, 
$\Lambda_{L_0}$
is
$(\gamma, E)$-regular}
\right)
\ge
1 - L_0^{-p}
\]
for some 
$\gamma > 0$, $p > 2d$
and some 
$L_0 > 0$
sufficiently large.
}\\

\noindent
$I$
can be taken in regions mentioned in the paragraph preceding  Definition in this section. 
This condition, together with Wegner's estimate, guarantee to apply the multiscale analysis, and from which the following fact is deduced
\cite{vonDK} : 
we can find 
$\alpha = \alpha(p,d)$, 
$1 < \alpha < 2$
such that, putting
\[
\Lambda_k (x) = \Lambda_{L_k} (x), 
\quad
L_k = L_{k-1}^{\alpha},
\quad
1 < \alpha < 2, 
\quad
x \in  {\bf Z}^d, 
\]
we have 
\begin{equation}
{\bf P}
\left\{
\mbox{For all 
$E \in I$
either 
$\Lambda_k(x)$
or
$\Lambda_k(y)$
is
$(\gamma, E)$-regular}
\right\}
\ge 
1 - L_k^{-2 p}
\label{MSA}
\end{equation}
for any 
$x, y$
with
$|x - y | > L_k$.
The first result
implies the distribution of localization centers are ``thin" in 
${\bf Z}^d$. 
\begin{theorem}
{\bf (localization centers are thin)}\\
Let
$d_k = | \Lambda_k |^{-1} k^{-2}$, 
$E \in I$, 
$J_k = (E - \frac {d_k}{2}, E + \frac {d_k}{2})
(\subset {\bf R})$, 
$k=1, 2, \cdots$.
Then for
a.e. 
$\omega$, 
we can find 
$k_0 = k_0 (\omega)$
such that, if
$k \ge k_0$, 
there are no eigenvalues of 
$H$
in 
$J_k$
localized in 
$\Lambda_k$.
\label{localized centers are thin}
\end{theorem}
\begin{remark}
Theorem 
\ref{localized centers are thin}
states that, for any 
$E \in I$, 
localization center run away from the origin, as the corresponding eigenvalue approaches 
$E$. 
It implies 
the distribution of the localization centers is thin,
while the eigenvalues are dense in 
$I$. 
A naive explanation of this fact is : 
we have infinite number of eigenvalues near 
$E$
while the number of states is proportional to the volume, 
by the finiteness of the density of states. 
\end{remark}
\begin{remark}
As the density of states obeys 
the Lifschitz tail asymptotics on  the bottom of the spectrum, another estimate is obtained there : 
$d_k$
can be replaced by 
$d_L = (a, a + \frac {1}{| \Lambda_L |^{2/d}})$, 
$a = \inf \sigma (H)$
(Simon \cite{Simon-Lifschitztail}).
\end{remark}
\begin{remark}
Theorem \ref{localized centers are thin}
holds true for random Hamiltonians 
where the multiscale analysis is applicable and Wegner's estimate holds. 
\end{remark}
\begin{theorem}
{\bf (localization centers are repulsive)}\\
Let 
$d_k = | \Lambda_k |^{-2}k^{-2}$, 
$k =1, 2, \cdots$.
For 
a.e. 
$\omega$, 
the following event occurs. 
For any 
$x \in {\bf Z}^d$, 
there exists 
$k_0 = k_0(\omega, x)$
such that for 
$k \ge k_0(\omega, x)$
and any interval 
$J \subset I$
with 
$|J| \le d_k$, 
there is at most one eigenvalue of 
$H$
in 
$J$
localized in 
$\Lambda_k (x)$. 
\label{localized centers are repulsive}
\end{theorem}
\begin{remark}
Theorem \ref{localized centers are repulsive}
implies : 
for a.e. $\omega$ and for any eigenvalue 
$E = E_j (\omega) \in I$
of 
$H$, 
we can find
$k_1 = k_1 (\omega,E_j(\omega))$
such that, for any 
$k \ge k_1$, 
we have no eigenvalues of 
$H$
in 
$J_k = (E_j(\omega) - \frac {d_k}{2}, 
E_j(\omega) + \frac {d_k}{2})$
localized in 
$\Lambda_k(x)$
for any 
$x \in X(E_j(\omega))$
except 
$E_j(\omega)$ 
itself.

Hence
this theorem roughly states that,  for a localization center 
$x$
with energy 
$E$, 
any other localization center must be away from 
$x$
at least in the distance of 
$L_k/2$, 
if the corresponding eigenvalues are within the distance of 
$| \Lambda_k |^{-2}$
from 
$E$. 
And 
this happens simultaneously for all eigenvalues in 
$I$
almost surely. 
\end{remark}
\begin{remark}
Theorem \ref{localized centers are repulsive}
also proves that eigenvalues in 
$I$
are simple almost surely. 
Indeed, this is done by Klein-Molchanov \cite{KM}
by the argument similar to ours, but without relying on the multiscale analysis. 
\end{remark}
\begin{remark}
One of the essential ingredients of the proof of 
Theorem \ref{localized centers are repulsive}
is the estimate of the probability to have more than two eigenvalues on a given interval obtained by Minami 
\cite[Lemma 2]{Minami}, 
which is also essential to prove the absence of repulsion of eigenvalues, 
and known to hold in the Anderson model only so far. 
Hence 
something different could be expected for the acoustic type operator, in which the level repulsion is known to occur
(Grenkova-Molchanov-Sudarev \cite{Grenkova}). 
\end{remark}
\begin{remark}
This result concerns 
the distribution of the localization centers which holds for almost surely. 
If one is interested in the fluctuation of those, 
by proceeding along the ideas in \cite{Minami}, 
one could expect the Poisson-type behavior also for localization centers, under a suitable scaling.
To verify this observation 
would be an interesting problem\footnote{The author would like to thank Rowan Killip to pointing this out. }
\cite{Nakano}. 
\end{remark}

In the following section, 
we prove
Theorem \ref{localized centers are thin}, 
\ref{localized centers are repulsive}, 
along the naive argument given at the beginning of this section, which is done by making use of the machinery developed by 
Germinet-De Bi\`evre,
 Damanik-Stollmann, 
 Klein-Molchanov and  
 Minami \cite{BG, DS, KM, Minami} : 
(i)
If we have 
an (resp. at least two) eigenvalue in an interval 
$J (\subset I)$, 
the corresponding eigenfunction is exponentially small outside a box 
$\Lambda$
surrounding its localization center
\cite{BG, DS}.
Then 
$H_{\Lambda}$
has an (resp. at least two) eigenvalue in 
$J$
\cite{KM}.
(ii)
We estimate the event that 
$H_{\Lambda}$
has an (resp. at least two) eigenvalue in 
$J$
by Wegner's (resp. Minami's) estimate.
Then 
the usual Borel-Cantelli argument gives the assertion.
In Appendix, 
we collect lemmas used in these proofs borrowed from 
\cite{BG, DS, KM, Minami}.
\section{Proof of Theorems}
We first set 
\begin{eqnarray*}
E_j &=& \Bigl\{
\omega \in \Omega : 
\mbox{
For some 
$E \in I$
and some
$x, y \in \Lambda_{3 L_{j+1}}$
with 
$\Lambda_j (x) \cap \Lambda_j (y) = \emptyset$, }
\\
&&\qquad\qquad
\mbox{
$\Lambda_j (x)$, $\Lambda_j(y)$
are both 
$(\gamma, E)$-singular}
\Bigr\}
\\
\Omega_k &=& \bigcap_{j \ge k} E_j^c.
\end{eqnarray*}
Then by (\ref{MSA}), 
\begin{equation}
{\bf P} (\Omega_k)
\ge
1 - C_1(\alpha, d, p) L_k^{2d \alpha - 2 p}
\label{Omega}
\end{equation}
for some 
$C_1= C_1(\alpha, d, p)$.
In what follows, 
we take and fix any 
$0 < \gamma' < \gamma$, 
and 
$k_0 (\alpha, d, \gamma)$, 
$k_1(\alpha, d, \gamma, \gamma')$, 
$L_0(\gamma')$, 
and 
$C_2$
are positive constants given  in Appendix.
\begin{lemma}
{\bf \mbox{}}\\
Let 
$J=(a,b)( \subset I)$
and let 
$\epsilon_{L_k} = C_2 e^{-\gamma' L_k/2}$. 
If 
$k \ge k_0(\alpha, d, \gamma) \vee k_1(\alpha, d, \gamma, \gamma')$, $L_k \ge L_0(\gamma')$, 
the following estimates hold.
\label{whole Hamiltonian}
\begin{eqnarray*}
(1) \quad &&
{\bf P}
\left(
\mbox{we have an eigenvalue of 
$H$ 
in 
$J$ localized in $\Lambda_k$}
\right)
\\
&& \qquad
\le
\| \rho \|_{\infty}
( | J |+ 2\epsilon_{L_k} )   | \Lambda_{3k} |
+
C_1(\alpha, d, p)
L_k^{2d \alpha -2 p}.
\\
(2) \quad &&
{\bf P}
\left(
\mbox{we have at least 2 eigenvalues of 
$H$ 
in 
$J$ 
localized in 
$\Lambda_k$}
\right)
\\
&& \qquad \le 
\pi^2 \| \rho \|^2_{\infty}
| \Lambda_{3 L_k} |^2
( | J | + 2\epsilon_{L_k} )^2
+
C_1(\alpha, d, p)
L_k^{2d \alpha -2 p}.
\end{eqnarray*}
\end{lemma}
\begin{proof}
(1)
Let 
\[
A_k := \left\{
\omega \in \Omega : 
\mbox{we have an eigenvalue of 
$H$ 
in 
$J$ localized in $\Lambda_k$}
\right\}.
\]
Let 
$\omega \in A_k \cap \Omega_k$.
Then we have an eigenvalue 
$E \in J$
localized in 
$\Lambda_k$
and since
$j \ge k_0 \vee k_1$, $L_k \ge L_0$, 
the corresponding eigenfunction 
$\phi$
satisfies
\[
\| ( 1- \chi_{3 L_k} ) \phi \|
\le
e^{- \gamma' L_k/2}
\]
by Lemma \ref{decay estimate2}.
Then 
the argument of the proof of 
Lemma \ref{approximation}
shows that 
$H_{\Lambda_{3 L_k}}$
has an eigenvalue in 
$(a - \epsilon_{L_k}, b + \epsilon_{L_k})$.
By 
Wegner's estimate:
\footnote{$\{ E_j (\Lambda) \}_{j=1}^{| \Lambda |}$
is the set of eigenvalues of 
$H_{\Lambda}$}
\[
{\bf P}(\sharp \{ E_j (\Lambda) \in J \} \ge 1)
\le
\| \rho \|_{\infty} | \Lambda | \cdot | J |,
\]
we have
$
{\bf P} (A_k \cap \Omega_k)
\le
\| \rho \|_{\infty}  | \Lambda_{ 3 L_k } | ( | J |+ 2\epsilon_{L_k} ).
$
\\
(2)
Let 
\[
B_k := \left\{ \omega \in \Omega : 
\mbox{we have at least 2 eigenvalues of 
$H$ 
in 
$J$ 
localized in 
$\Lambda_k$}
\right\}
\]
By the same argument 
in the proof of Lemma \ref{whole Hamiltonian}(1), 
if
$\omega \in B_k \cap \Omega_k$, 
$H_{\Lambda_{3 L_k}}$
has at least two eigenvalues in 
$(a - \epsilon_{L_k}, b + \epsilon_{L_k})$.
By Minami's estimate
\cite[Lemma 2]{Minami}, 
\cite[Appendix]{KM}:
\[
{\bf P}\left(
\{ \sharp \{ E_j (\Lambda) \in J \} \ge 2 \}
\right)
\le
\pi^2 \| \rho \|^2_{\infty} | \Lambda |^2 | J |^2, 
\]
we have
$
{\bf P} (B_k \cap \Omega_k)
\le
\pi^2 \| \rho \|^2_{\infty}
\| \Lambda_{3 L_k} |^2 ( | J |+ 2\epsilon_{L_k} )^2.
$
\QED
\end{proof}
\noindent
{\it Proof of Theorem \ref{localized centers are thin}}
\\
We consider the following event. 
\[
A_k = \left\{
\omega \in \Omega : 
\mbox{ 
We have an eigenvalue of $H$ in $J_k$
localized in $\Lambda_k$
} 
\right\}.
\]
By Lemma \ref{whole Hamiltonian}, 
if 
$k \ge k_0(\alpha, d, \gamma) \vee k_1(\alpha, d, \gamma, \gamma')$
and
$L_k \ge L_0(\gamma')$, 
we have
\[
{\bf P}(A_k)
\le
\| \rho \|_{\infty} | \Lambda_{3 L_k} | 
 ( d_k + 2\epsilon_{ L_k} )
+
C_1 (\alpha, d, p) L_k^{2d(\alpha-1) - 2 p}
\]
Since
$d_k +2 \epsilon_{L_k} \le 2 d_k$
for sufficiently large 
$k$, 
$\sum_k {\bf P} (A_k) < \infty$. 
The
Borel-Cantelli
argument then proves the assertion of 
Theorem \ref{localized centers are thin}. 
\QED\\

\noindent
{\it Proof of Theorem \ref{localized centers are repulsive}}
\\
We consider the following events. 
\begin{eqnarray*}
B_k (x, J)
&=&
\left\{
\omega\in \Omega : 
\mbox{at least two eigenvalues of $H$ in $J$
 localized in $\Lambda_k(x)$}
\right\}, 
\\
B_k (x)
&=&
\Bigl\{
\omega\in \Omega : 
\mbox{at least two eigenvalues of $H$ in $J$}
\\
&&\qquad\qquad
\mbox{ 
for some 
$J(\subset I)$ with 
$|J| \le d_k$ in $\Lambda_k(x)$}
\Bigr\}.
\end{eqnarray*}
Suppose
$| J | \le 2 d_k$, $J \subset I$. 
Then
by the argument in the proof of 
Lemma \ref{whole Hamiltonian}(2), 
\[
{\bf P}(B_k(x, J) \cap \Omega_k)
\le
\pi^2 \| \rho \|_{\infty}^2 ( 3 d_k )^2
\cdot
| \Lambda_{3 L_k} |^2
\]
for 
$k$
sufficiently large. 
Here we use the argument in 
\cite[Lemma 2]{KM}
and cover the interval 
$I$
by those 
$J(i, d_k)$, $i = 1, 2, \cdots, N_k$
of width 
$2 d_k$
such that the left end of 
$J(i+1, d_k)$
coincides with the mid point of 
$J(i, d_k)$. 
Then 
$N_k \le \frac {| I |}{2 d_k} \cdot 2 = \frac {| I |}{d_k}$
and any interval (in $I$) of width less than 
$d_k$
is contained by some 
$J(i, d_k)$.
Hence
\[
{\bf P} (B_k (x) \cap \Omega_k)
\le
\pi^2 \| \rho \|_{\infty}^2 
( 3 d_k )^2 \cdot \frac {|I|}{d_k}
\cdot
| \Lambda_{3 L_k} |^2
\]
for large 
$k$
and therefore 
$\sum_k {\bf P}(B_k(x)) < \infty$
by (\ref{Omega}).
By the Borel-Cantelli lemma, 
$\Omega(x) = \liminf_{k \to \infty} B_k^c(x)$
satisfies
${\bf P}(\Omega(x)) = 1$
and for 
$\omega \in \Omega(x)$
we can find
$k_0 = k_0(\omega,x)$
such that for any
$k \ge k_0(\omega, x)$
and for any interval 
$J (\subset I)$ 
with 
$| J | \le d_k$, 
we have at most one eigenvalue of 
$H$
in 
$J$
localized in 
$\Lambda_k (x)$. 
For 
$\omega \in \Omega' = \bigcap_{x\in {\bf Z}^d} \Omega(x)$, 
the event described in the statement of 
Theorem \ref{localized centers are repulsive} occurs. 
\QED

\section{Appendix}
In this section, 
we state Lemmas used in section 2, which are borrowed from 
\cite{BG, DS, KM, Minami}. 
The following lemma is 
\cite[Lemma 3.5]{BG}.
\begin{lemma}
{\bf (\cite[Lemma 3.5]{BG})}\\
We can find a constant 
$k_0 = k_0 (\alpha, d, \gamma)$
such that, if 
$k \ge k_0$
and 
$\phi \in l^2 ({\bf Z}^d)$
satisfies
$H \phi = E \phi$, 
then 
$\Lambda_{L_k} (x_{\phi})$
is
$(\gamma, E)$-singular.
\label{localized centers live in bad box}
\end{lemma}
In what follows, we take and fix any 
$\gamma'$ 
with 
$0 < \gamma' < \gamma$.
\begin{lemma}
{\bf \mbox{}}\\
We can find a constant 
$k_1 = k_1 (\alpha, d, \gamma, \gamma')$ 
such that, if 
$k\ge k_0(\alpha, d, \gamma) \vee k_1 (\alpha, d, \gamma, \gamma')$, 
$\omega \in \Omega_k$
and if 
$\phi \in l^2({\bf Z}^d)$
satisfies 
$H \phi = E \phi$, $\| \phi \| = 1$
and localized in 
$\Lambda_k$, 
\label{decay estimate2}
\[
\| ( 1 - \chi_{\Lambda_{3 L_k}}) \phi \|
\le
e^{- \gamma' L_k/2}, \quad \gamma' < \gamma.
\]
\end{lemma}
Lemma \ref{decay estimate2}
is proved along the argument in 
\cite[Step 3, Theorem 3.1]{DS}.\\

\noindent
{\it Sketch of proof }
We divide
$\Lambda_{3 L_k}^c$
into annulus : 
$
\Lambda_{3 L_k}^c 
=
\bigcup_{i \ge k} M_i, 
\;
M_i = \Lambda_{3 L_{i+1}} \setminus \Lambda_{3 L_i}, 
\;
i \ge k.
$
Then we have
$
\| (1 - \chi_{3 L_k}) \phi \|^2
\le 
\sum_{i = k}^{\infty}
\| \chi_{M_i} \phi \|^2
\le 
\sum_{ i = k }^{\infty} \sum_{x \in M_i}
| \phi (x) |^2.
$
Since 
$\Lambda_{L_i} (x_{\phi}) \cap \Lambda_{L_i}(x) = \emptyset$
for any 
$x \in M_i$, 
$\Lambda_{L_i}(x)$
is 
$(\gamma, E)$-regular
by Lemma \ref{localized centers live in bad box}.
\QED
\begin{lemma}
{\bf \mbox{}}\\
We can find positive constants
$C_2$, 
$L_0 = L_0 (\gamma')$
with the following property. 
If 
$\varphi_1, \varphi_2$
satisfy
\begin{eqnarray*}
&&
H \varphi_j = E_j \varphi_j, 
\quad
\| \varphi_j \| = 1, 
\quad
E_j \in J = (a,b), 
\quad
j=1,2,
\\
&&
\| ( 1 - \chi_{\Lambda_L}) \varphi_j \|
\le
e^{ - \gamma' L/2}, 
\quad
j = 1,2, 
\quad
L \ge L_0, 
\end{eqnarray*}
then 
$H_{\Lambda_L}$
has at least two eigenvalues in 
$(a - \epsilon_L, b + \epsilon_L)$
where
$\epsilon_L = C_2 e^{-\gamma' L/2}$.
\label{approximation}
\end{lemma}
The proof 
is found in \cite{KM} : 
we orthonormalize
$\varphi_j^{\Lambda} = \chi_{\Lambda} \varphi_j$, 
$j=1,2$
and estimate from below the trace of the spectral projection of 
$H$
corresponding to the interval 
$(a- \epsilon_L, b+\epsilon_L)$.\\

\noindent {\bf Acknowledgement }
The author would like to thank Professors 
Abel Klein, Rowan Killip, Nariyuki Minami and a referee 
for their discussions and comments. 

%

\end{document}